\documentclass[sigconf,natbib]{acmart}
\usepackage[skip=0pt]{caption}
\usepackage{color}
\usepackage{multirow}
\usepackage{dsfont}
\usepackage{relsize}
\usepackage{colortbl}
\usepackage{algorithm,algorithmic}
\usepackage{xcolor}
\usepackage{mathtools}

\definecolor{silver}{rgb}{0.753,0.753,0.753}

\usepackage{comment}
\usepackage{amsmath}
\usepackage{accents}
\usepackage{hyperref}
\usepackage{enumerate}
\usepackage[shortlabels,inline]{enumitem}
\usepackage{subcaption}

\definecolor{ballblue}{rgb}{0.13, 0.67, 0.8}
\definecolor{caribbeangreen}{rgb}{0.0, 0.8, 0.6}
\definecolor{cherryblossompink}{rgb}{1.0, 0.72, 0.77}
\definecolor{fluorescentorange}{rgb}{1.0, 0.75, 0.0}
\definecolor{skyblue}{rgb}{0.53, 0.81, 0.92}
\definecolor{stildegrainyellow}{rgb}{0.98, 0.85, 0.37}
\AtBeginDocument{%
 \providecommand\BibTeX{{%
  \normalfont B\kern-0.5em{\scshape i\kern-0.25em b}\kern-0.8em\TeX}}}
  
\copyrightyear{2023}
\acmYear{2023}
\setcopyright{rightsretained}
\acmConference[XXX '23]{Proceedings of the XXX}{XXX}{XX, XXX}

\acmDOI{}
\acmISBN{}
\begin{document}
\title{Enhancing Documents with Multidimensional Relevance Statements in Cross-encoder Re-ranking}

\author{Rishabh Upadhyay}
\authornotemark[1]
\email{r.upadhyay@camus.unimib.it}
\orcid{0000-0001-9937-6494}
\affiliation{%
  \institution{Department of Informatics, Systems, and Communication \\University of Milano-Bicocca}
  \streetaddress{Viale Sarca, 336 -- 20126}
  \city{Milan}
  \country{Italy}
}

\author{Arian Askari}
\email{a.askari@liacs.leidenuniv.nl}
\affiliation{%
  \institution{Leiden Institute of Advanced Computer Science, \\Leiden University}
  \city{Leiden}
  \state{Ohio}
  \country{Netherlands}
}

\author{Gabriella Pasi}
\email{gabriella.pasi@unimib.it}
\orcid{0000-0002-6080-8170}
\affiliation{%
  \institution{Department of Informatics, Systems, and Communication \\University of Milano-Bicocca}
  \streetaddress{Viale Sarca, 336 -- 20126}
  \city{Milan}
  \country{Italy}
}

\author{Marco Viviani}
\email{marco.viviani@unimib.it}
\orcid{0000-0002-2274-9050}
\affiliation{%
  \institution{Department of Informatics, Systems, and Communication \\University of Milano-Bicocca}
  \streetaddress{Viale Sarca, 336 -- 20126}
  \city{Milan}
  \country{Italy}
}

\begin{abstract}
In this paper, we propose a novel approach to consider multiple dimensions of relevance beyond topicality in cross-encoder re-ranking. On the one hand, current multidimensional retrieval models often use naïve solutions at the re-ranking stage to aggregate multiple relevance scores into an overall one. On the other hand, cross-encoder re-rankers are effective in considering topicality but are not designed to straightforwardly account for other relevance dimensions. To overcome these issues, we envisage enhancing the candidate documents -- which are retrieved by a first-stage lexical retrieval model -- with ``relevance statements'' related to additional dimensions of relevance and then performing a re-ranking on them with cross-encoders. In particular, here we consider an additional relevance dimension beyond topicality, which is credibility. We test the effectiveness of our solution in the context of the Consumer Health Search task, considering publicly available datasets. %both TREC Health Misinformation 2020 and CLEF e-Health 2020 datasets. 
Our results show that the proposed approach statistically outperforms both aggregation-based and cross-encoder re-rankers.

% 1. what did we do
% 2. why did we do it
% 3. How did we do it
% 4. What did we find
% 5. What do we think it means?
\end{abstract}
\maketitle
\section{Introduction}
\label{sec:introduction}
In recent years, there has been an increasing interest in addressing the problem of implementing effective retrieval models that consider different dimensions of relevance across various domains and tasks in the field of Information Retrieval \cite{da2012multidimensional,van2017concept, Goeuriot2021CLEFEE,Trec2020}. Today, document ranking is often achieved by performing a \emph{first-stage retrieval}, usually focused on topical relevance, to efficiently identify a subset of relevant documents from the entire collection; on this subset, a \emph{re-ranking} stage is performed, where topicality and/or additional dimensions of relevance may be considered.

BM25 \cite{robertson1994some} is often used as a first-stage retrieval model because of its effectiveness and efficiency. Concerning re-ranking models, some are based on the \emph{aggregation} of the topicality score obtained from the first-stage retrieval model and other relevance scores related to additional relevance dimensions \cite{da2009multidimensional,da2009prioritized}. In doing so, simple and compensatory aggregation techniques are used in most cases \cite{fox1993combination}. Other re-rankers exploit the potential of \emph{cross-encoders} \cite{devlin2018bert}. 
A recent study, in particular, has improved the effectiveness of such kind of re-ranker by injecting the BM25 score obtained in the first-stage retrieval as an input token to the cross-encoder re-ranker \cite{askari2023injecting}. However, despite the effectiveness of this type of approach, the limitation of considering only topical relevance remains.

Therefore, in this paper, we aim to explore the impact of incorporating other dimensions of relevance into a cross-encoder for document re-ranking. In particular, instead of manipulating the input structure of the cross-encoder with an additional \emph{relevance score} for an additional relevance dimension, we integrate a so-called \emph{relevance statement} into the document. This statement is constituted by a text related to the relevance dimension under consideration and its associated relevance score. This ``enhanced'' document is provided, along with the query, as input to the cross-encoder to obtain the final relevance score.

To illustrate and assess the proposed model we consider, in particular, the \emph{Consumer Health Search} (CHS) task \cite{goeuriot2020overview}. As an additional dimension of relevance beyond topicality, we take into account the genuineness of the information, which in many IR tasks so far has been referred to as \emph{credibility} \cite{viviani2017credibility}.
%By our extensive analysis, we show that the $CE_{\textrm{\emph{rel.stat}}}$ model leverages both the content of the document and external information, such as credibility, to compute the relevance score with the query.
% .in order to analyse and investigate how this challenge (injecting non-topicality score) should be addressed, we addressed the following research question:
% \par
% \textbf{(RQ1)} How effective is it to inject a task-specific score such as credibility score instead of a topicality score? 
% % \par \noindent We set up x in order to answer RQ1... Next ...:
% \par
% \textbf{(RQ2)} What is the effectiveness of CE\textsubscript{StatementCAT} compared to common approached and CE\textsubscript{CredCAT} for combining the credibility score with cross-encoders?
% % \par \noindent
% % We investigate x in order to answer RQ3... Next ...:
% \par
% \textbf{(RQ3)} To what extent the CE\textsubscript{CredCAT} attributes to the added task specific statement?
% % \par \noindent
% % To doing so, we ...:
% \par
%
To the best of our knowledge, there is no prior work studying the effectiveness of cross-encoder re-rankers by considering non-topical relevance dimensions by document enhancement. Hence, the contributions of our work are:
\begin{itemize}
    \item We propose the $CE_{\textrm{\emph{rel.stat}}}$ model, whose purpose is to exploit the potential of cross-encoders in the re-ranking phase by exploiting the multidimensional nature of relevance, and to overcome the compensatory effect of current aggregation-based multidimensional relevance models; %it integrates a \textcolor{magenta}{relevance score and a relevance statement} in the document \textcolor{magenta}{for each considered relevance dimension (in our case, credibility), and provides this enriched document, in addition to the query, as input to the cross-encoder}. This enables the model \textcolor{magenta}{to better account for} the multidimensional relevance of documents, by taking into account both the topicality and credibility dimensions.
    \item We analyze the effectiveness of the proposed model compared to state-of-the-art %models 
    and standard baselines by conducting extensive experiments on the TREC-2020 Health Misinformation and CLEF-2020 eHealth datasets \cite{Trec2020,goeuriot2020overview};
    %A series of experiments to evaluate the effectiveness of the proposed model compared to several other retrieval models,including BM25, BERT re-rankers, and other input manipulation cross-encoder-based models on TREC-2020 Health Misinformation and CLEF-2020 eHealth Datasets.
    \item We provide a qualitative analysis for explainability, showing the importance of including relevance statements in documents.
    % We analysed the input qualitatively using Shapley values (SHAP). By applying the SHAP method to the $CE_{\textrm{\emph{rel.stat}}}$ model, we gained insights into the importance of the different components of the input, including the credibility score and the textual information in the document.
\end{itemize}
%After a discussion of related work in Section 2, we present our proposed method, the $CE_{\textrm{\emph{rel.stat}}}$ model, in Section 3 that integrates credibility score with textual information\todo[color=blue!40]{Extra para?} in the document to enhance multidimensional document retrieval. We describe the experimental setup, including the dataset, baselines, and implementation details, in Section 4. In Section 5, we present the evaluation results of the proposed model and analyse its effectiveness. Finally, in Section 6, we conclude the paper and discuss the future works.

\section{Related work}\label{sec:related_work}
% \textbf{Modifying input of Cross-encoders.}

Among the various multidimensional retrieval approaches that apply re-ranking based on the \emph{aggregation} of the topicality score to other relevance dimension scores -- such as credibility, correctness, and understandability -- are \cite{fernandez2020citius,pradeep2020h2oloo, abualsaud2021uwaterloomds,bondarenko2021webis,zhang2022ds4dh,schlicht2021upv,upadhyay2022unsupervised}. These approaches use different solutions to calculate these relevance scores, but all use approaches to aggregate these scores based on simple \emph{linear aggregation} \cite{fernandez2020citius,pradeep2020h2oloo, abualsaud2021uwaterloomds,bondarenko2021webis,upadhyay2022unsupervised}, and \emph{rank fusion} methods \cite{zhang2022ds4dh,schlicht2021upv}. These methods have a compensatory effect, which is difficult to explain, between the different dimensions of relevance.

In approaches that use cross-encoders for re-ranking, two so-called \emph{sequences} -- i.e., the \emph{query} $q$ and a candidate \emph{document} $d$ -- are concatenated and fed into a \emph{Transformer} model like BERT \cite{devlin2018bert}. Thus, Transformer \emph{attention heads} can directly model which elements of one sequence are correlated with elements of the other, allowing a (topical) relevance score $\sigma$ to be calculated. Formally:
\begin{equation}
\label{eq:bert}
    \sigma(q, d) = \mathrm{CE}(\mathrm{[CLS]} \ q \ \mathrm{[SEP]} \ d \ \mathrm{[SEP]}) \cdot W
    % \textrm{\emph{Input Sequence}} = [[CLS]\tilde{q}[SEP]\tilde{d}[SEP]] 
\end{equation}
where CE is the cross-encoder, CLS and SEP are special tokens to represent the classifier token and the separator token, and $W$ is a learned matrix that represents the relationship between the query and document representations. In this context, the recent $CE_{\textrm{BM25CAT}}$ model \cite{askari2023injecting} has been proposed to improve the effectiveness of BERT-based re-rankers by injecting the topicality score obtained by a first-stage BM25 model as a token (BM25) into the input of the cross-encoder. Formally: 
\begin{equation}
\label{eq:arian}
    \sigma(q, d) = \mathrm{CE}(\mathrm{[CLS]} \ q \ \mathrm{[SEP]} \ \mathrm{BM25} \ \mathrm{[SEP]} \ d \ \mathrm{[SEP]}) \cdot W
    % \textrm{\emph{Input Sequence}} = [[CLS]\tilde{q}[SEP]\tilde{d}[SEP]] 
\end{equation}
However, this approach does not focus on multidimensional relevance. %, \textcolor{magenta}{and it simply considers the addition of a BM25 score to the input structure of the encoder, not contextual information in the form of a statement in the document. 
Other approaches that modify the input to more effective retrieval are %and only adds a BM25 token as a signal of topical relevance into the input; (2) they add one token into the input, while our proposed method adds a textual statement, and we show that our method is statistically significantly effective in multi-dimensional retrieval. Boualili et al. Their motivation for injecting a numerical token was related to the fact that BERT-based re-rankers could process and work with numbers \cite{wallace2019nlp,thawani2021representing} while they struggle with general common-sense reasoning.
\cite{boualili2020markedbert,boualili2022highlighting}; they propose a method for highlighting exact matching signals by marking the start and the end of each occurrence of the query terms by adding markers to the input. In \cite{al2022arabglossbert}, the authors experimented with the inclusion of various supervised signals into the input of the cross-encoder to emphasize target words in context, while in \cite{li2022markbert} the authors injected boundary markers between contiguous words for Chinese named entity identification.
% \textbf{Mutli-dimensional retrieval with cross-encoders. }
%Re-ranking using other relevance dimensions, such as credibility, correctness, and understandability, has been incorporated mainly using linear aggregation \cite{fernandez2020citius,pradeep2020h2oloo, abualsaud2021uwaterloomds,bondarenko2021webis}, and rank fusion methods\cite{zhang2022ds4dh,schlicht2021upv}. 
However, also these studies do not account for multidimensional relevance. %in the context of cross-encoder re-rankers. %on multi-dimensional retrieval with textual dimensions. The recent but less relevant work, as they focus on other dimensions such as audio and image, are \cite{juliao2020exploring,zhou2022vlue}.
% \citet{xue2021hierarchical} propose a ..
% \begin{figure*}[!t]
%    \begin{minipage}{0.48\textwidth}
%      \centering
%      \includegraphics[width=1.0\linewidth]{figures/ce.pdf}
%      \caption{Regular cross-encoder input}\label{fig:ce}
%    \end{minipage}\hfill
%    \begin{minipage}{0.48\textwidth}
%      \centering
%      \includegraphics[width=1.0\linewidth]{figures/cerelstat.pdf}
%      \caption{Enriching document in input}\label{fig:proposed_method}
%    \end{minipage}
% \end{figure*}

\section{The $\textrm{\emph{CE}}_{\textrm{\emph{rel.stat}}}$ Model for Re-ranking} 
\label{sec:ce_relstat_model}%baseline: bm25_cat, cred_cat, multi_cat: bm25_cat [sep] cred_cat, statement_cat: score injected as statement (ours)

\begin{figure}[!h]
  \centering
\includegraphics[width=\linewidth]{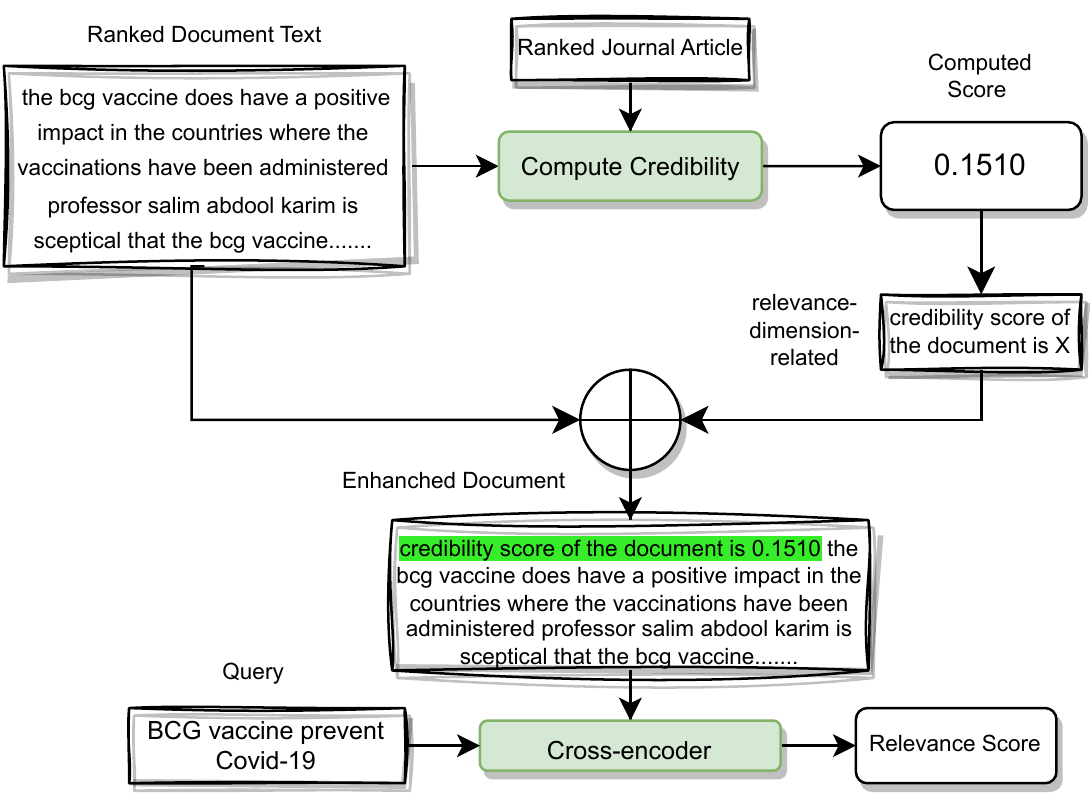}
\caption{The $CE_{\textrm{\emph{rel.stat}}}$ re-ranker.}\label{fig:architecture}
\end{figure}

%To address the challenge of retrieval tasks involving multiple dimensions, we propose an approach that integrates task-specific information and scores into the document prior to feeding it into the cross-encoder. This approach is demonstrated and evaluated for the retrieval of health-related documents that are both topically relevant and credible. 
The cross-encoder-based model proposed in this paper to perform re-ranking, namely $CE_{\textrm{\emph{rel.stat}}}$, is based on performing four steps, i.e., $(i)$ an initial \emph{retrieval} phase by using BM25, $(ii)$ the \emph{computation} of a \emph{relevance score} for an additional relevance dimension, $(iii)$ the \emph{enhancement} of the retrieved documents with a text related to the additional relevance dimension in the form of a
%relevance-dimension related textual information in the form of a
\emph{relevance statement}, and $(iv)$ the actual \emph{re-ranking} that occurs by feeding the cross-encoder with the \emph{query} and the related \emph{enhanced documents}. Figure \ref{fig:architecture} illustrates these steps in the context of the CHS task, considering \emph{credibility} as an additional dimension of relevance.

The steps shown in  Figure 1 are detailed below, also instantiated with respect to CHS and credibility.

\subsection{First-stage Retrieval: BM25}
The first-stage retrieval in our framework is based on the BM25 model \cite{robertson2009probabilistic}. BM25 calculates a \emph{topicality score} based on word frequency and distribution in the query and document, providing a list of the most relevant documents. It is effective and efficient, making it a popular choice as the first-stage ranker in Information Retrieval Systems \cite{anand2021serverless,kamphuis2020bm25,gao2021complement}.

\subsection{Calculation of the Relevance Score}\label{sec:cred}

Depending on which dimension(s) of relevance to consider in addition to topicality, different methods can be taken into account to obtain a relevance score for each dimension, based on the nature of the relevance dimension(s). In our case, to calculate the \emph{credibility score}, we employed a recent state-of-the-art approach \cite{upadhyay2022unsupervised}. It is an unsupervised solution that overcomes the problem of dealing with labeled datasets and obtained high effectiveness on the CHS datasets that are used in our work for evaluation purposes. This approach involves comparing the content of retrieved documents, given a query, with scientific articles, which are considered reliable sources of evidence for the same query. Both the documents and scientific articles are represented using BioBERT %, which was chosen over BERT for this study due to its better effectiveness in the health domain 
\cite{lee2020biobert}; to calculate the credibility score of each retrieved document $d$ with respect to a query $q$, denoted as $\textrm{\emph{cred}}(d,q)$, a linear combination of the cosine similarity scores between $d$ and the top-$k$ scientific articles $j_j$ that were deemed relevant to the same query for which $d$ was retrieved, is performed. Formally: %This was done by taking a weighted average of the similarity scores:
\begin{equation}\label{eq:igs}
    \textrm{\emph{cred}}(d,q) = \omega_1 \cdot \cos(d,j_1) + \omega_2 \cdot \cos(d,j_2) + \ldots + \omega_k \cdot \cos(d,j_k)
\end{equation}
where $\omega_{1}, \omega_{2}, \ldots, \omega_{k}$, such that $\sum \omega_i=1$, and $\omega_{i}$ $\geq$ $\omega_{i+1}$ ($1 \leq i \leq k-1$). These weights allow assigning greater emphasis to the similarity scores according to %the article estimated to be most relevant to the query; indeed, each weight is dependent on 
the \emph{rank} of the retrieved articles $j_j$, as illustrated in detail in \cite{upadhyay2022unsupervised}.

\subsection{Document Enhancement}
In this phase, we enhance each document retrieved in the first-stage phase with a so-called \emph{relevance statement} that is related to the additional relevance dimension considered. This statement consists of the \emph{relevance score} obtained in the second phase plus a text related to the
additional relevance dimension associated with the score. %In our case, the enrichment process involves adding both the credibility score -- Equation (\ref{eq:igs}) -- and a \emph{credibility statement}. %This score could be generated using various methods, but we considered using the method explained in section \ref{cred}. 
Of course, this textual information should vary depending on the size of the relevance being considered. For example, if considering \emph{credibility}, the form of the statement can be:
%they can refer to distinct relevance dimensions, including topicality statements such as "\textit{topicality score is X}", credibility statements such as 
"\textit{credibility score X}", where $X$ is the relevance score associated with credibility. %, etc. They can also be constituted by combined statements including distinct relevance scores, such as "\textit{topicality score of the document is X and credibility score of the document is Y, \ldots}". 
In this case, being $d$ = "\emph{the bcg vaccine does have a positive \ldots}" the original document, its enhanced version becomes: $\tilde{d}$ = "\emph{STAT the bcg vaccine does have a positive \ldots}", where \emph{STAT} represents the considered statement. The statements actually tested in this work to enhance the documents will be illustrated in the section devoted to experimental evaluations.

% The values of \textit{X} and \textit{Y} \textcolor{magenta}{are those obtained by BM25 and Equation (\ref{eq:igs}) respectively, normalized in the same interval.}
% \begin{equation}
% \footnotesize
%     d'=STAT\_X\ the\ bcg\ vaccine\ does\ have\ a\ positive.....
% \end{equation}
% where $STAT\_X$ is the credibility-specific statement with score representation, and d' is the enriched document.
% % Experimental results with different representation are shown in section \ref{sec:result}.

\subsection{Cross-encoder Re-ranking}
% The \emph{ranking model}, denoted by ${K(q, d)}$, calculates the relevance score between a user query \emph{q} and a document \emph{d} from a large-scale collection $D$ (i.e., $d \in D$). This model is used downstream to an efficient retriever, such as BM25, in a retrieval and re-ranking pipeline. 
In the re-ranking phase, %we leverage our document enhancement technique that incorporates contextual information (statement) with relevance score (in this case credibility). This 
each enhanced document is fed to the cross-encoder for the computation of the relevance score between the document and the query. By enhancing the document with a relevance statement built as illustrated before, we provide the cross-encoder with an additional relevance dimension (such as credibility) beyond topicality, which is essential in multidimensional retrieval. Formally: %This approach addresses the challenge of incorporating other relevance dimension in information retrieval systems and can lead to improvement. For enriched document d' and query q, it is given as:
\begin{equation}
\label{eq:CErel}
    \sigma(q, \tilde{d}) = \mathrm{CE}(\mathrm{[CLS]} \ q \ \mathrm{[SEP]} \ \tilde{d} \ \mathrm{[SEP]}) \cdot W 
\end{equation}

\section{Experimental Evaluations} %pytorch, huggingface, sentence-transformers cross-encoder package etc
\label{sec:experiments}

%\subsection{The Consumer Health Search Task}
%\label{sec:taskdata}
For evaluation purposes, we focused on the \emph{ad-hoc retrieval} task within the TREC-2020 Health Misinformation Track \cite{Trec2020} and CLEF-2020 eHealth Track \cite{goeuriot2020overview}. Both tracks relate to \emph{Consumer Health Search} and consider \emph{credibility} an important relevance dimension in addition to topicality. A subset of 1 million documents from each track was used, % due to computational expenses, 
with the TREC-2020 Track covering 46 topics related to Coronavirus and the CLEF-2020 Track covering 50 medical condition topics.
%The TREC-2020 track had 46 topics related to coronavirus, each with a title, description, and answer that included an evidence link, while the CLEF-2020 track had 50 topics covering various medical conditions, each with a title only. 
%Since there were no predefined train, development, and evaluation split, we fine-tuned the proposed model using queries from one dataset and evaluated it on queries from the other dataset and vice versa. 
The TREC-2020 Health Misinformation Track provides binary labels for documents, with those meeting the criteria of being "topical and credible" labeled as "1" and the rest labeled as "0". As for the CLEF-2020 eHealth Track, the same binary labeling applies for both topicality and credibility. %, as no binary labels are provided. 
%We reported the evaluation metrics NDCG@10, P@10, MRR@10, and MAP.  The above MRR, NDCG, MAP, and P refer to Mean Reciprocal Rank, Normalized Cumulative Discount Gain, Mean Average Precision and Precision, respectively.

\subsection{Implementation Details} \label{sec:implementation}
We employed PyTerrier \cite{pyterrier2020ictir} for indexing and implementing the BM25 model. We created two indexes, one for TREC-2020 and another for CLEF-2020. As the considered document set is health-related, we used BioBERT ($dmis-lab/biobert-v1.1$\footnote{\url{https://huggingface.co/dmis-lab/biobert-v1.1}}) \cite{lee2020biobert} along with the base version of the BERT model for cross-encoder re-ranking training and inference. It should be noted that the tokenizer's vocabulary of BERT and BioBERT already includes integer and floating numbers for appropriate tokenization.
\par
Since no split was provided for TREC-2020 and CLEF-2020, we trained the cross-encoder on 80\% of the queries from one dataset (e.g. TREC-2020) and used the other query set as the development set, and selected all queries and documents from the other dataset (CLEF-2020) as the evaluation set and vice versa. We fine-tuned both the BERT and BioBERT models with a batch size of 4 and 512 tokens as the maximum sequence length for 10 epochs using the Adam optimizer with an initial learning rate set to $2 \times 10^{-5}$. We employed the Huggingface library \cite{wolf2019huggingface}, the Cross-encoder package of Sentence-transformers library \cite{reimers-2019-sentence-bert}, and PyTorch \cite{paszke2019pytorch} for training and inferencing.

\subsection{Baselines}\label{sec:base}
The baseline models considered to evaluate the proposed approach comparatively include: 
\begin{itemize}
    \item BM25: the BM25 model implemented by PyTerrier;
    \item \emph{WAM}: the aggregation-based multidimensional relevance model presented in \cite{upadhyay2022unsupervised}, based on a simple weighted average of distinct relevance scores. Specifically, weights associated with topicality and credibility are set as in the best model described in \cite{upadhyay2022unsupervised};%Topicality receives a weight of 0.4 and credibility a weight of 0.6;
    \item \emph{CE}: the original cross-encoder model for re-ranking \cite{devlin2018bert}, based on Equation (\ref{eq:bert});
    % \begin{equation}
    %     \sigma(q,d)=???
    % \end{equation}
    \item \textit{$CE_{BM25CAT}$}: the cross-encoder re-ranker \cite{askari2023injecting}, where normalized BM25 score is injected into the input structure of the cross-encoder, based on Equation (\ref{eq:arian});
    % \begin{equation}
    % \sigma(q, d) = \mathrm{CE}(\mathrm{[CLS]} \ q \ \mathrm{[SEP]} \ \mathrm{BM25} \ \mathrm{[SEP]} \ d \ \mathrm{[SEP]}) \ast W
    % \end{equation}
    \item \textit{$CE_{CredCAT}$}: a cross-encoder re-ranker, where a credibility score is injected into the input structure of the cross-encoder instead of the BM25 score 
    Formally:
    \begin{equation}
    \sigma(q, d) = \mathrm{\mathrm{CE}}(\mathrm{[CLS]} \ q \ \mathrm{[SEP]} \ \textrm{\emph{cred}} \ \mathrm{[SEP]} \ d \ \mathrm{[SEP]}) \cdot W
    \end{equation}
    where \emph{cred} is the credibility score;
    \item \textit{$CE_{BM25CredCAT}$}: a cross-encoder re-ranker, where both BM25 and credibility scores are injected into the input structure of the cross-encoder. Formally:
    \begin{equation}
    \sigma(q,d) = \mathrm{CE}(\mathrm{[CLS]} \ q \ \mathrm{[SEP]} \ \mathrm{BM25} \ \mathrm{[SEP]} \ \mathrm{\emph{cred}} \ \mathrm{[SEP]}\ d \ \mathrm{[SEP]}) \cdot W
    \end{equation}
\end{itemize}

\subsection{Results}\label{sec:result}

This section provides the experimental results of the proposed $\textrm{\emph{CE}}_{\textrm{\emph{rel.stat}}}$ model compared with the baselines, using \emph{Normalized Discounted Cumulative Gain} at 10 (NDCG@10), \emph{Precision} at 10 (P@10), \emph{Mean Reciprocal Rank} at 10 (MRR@10) and \emph{Mean Average Precision} (MAP) as evaluation metrics. All results are statistically significant according to a paired t-test $(p < 0.05)$ with Bonferroni correction for multiple testing \cite{weisstein2004bonferroni}. Specifically, before providing the final results, research questions had to be answered to optimize the parameters to be used in the proposed model.

%Previous studies have shown that BERT can process a specific range of numbers better \cite{wallace2019nlp}. Therefore, in order to enrich the document with the proper representation of the credibility score, we first addressed the following research question.

\paragraph{RQ1: What is the best representation of the credibility score for cross-encoder re-ranking?}
Previous studies have shown that BERT and other NLP model effectiveness can depend on the way in which numbers are encoded
%the choice of the number representation \textcolor{blue}{what do you mean here? the way in which numbers are represented
\cite{wallace2019nlp}. Hence, %Credibility scores were computed using the method and the weights used for linear aggregation as described in Section \ref{sec:cred}. Since credibility scores range from 0 to 1, we did not normalize them. We used 
we tested three types of representations for the credibility score, i.e., \emph{decimal} (dec.), \emph{integer} (int), and \emph{segmented} (seg.), i.e., each digit of the score is taken individually. To evaluate the effectiveness of such numerical representations for ranking purposes, we considered as a baseline the $CE$ cross-encoder model with BioBERT textual representation, and a simplified model compared to ours, which we call $\textrm{\emph{CE}}_{\textrm{\emph{rel.score}}}$, which enhances documents with just \emph{credibility scores}, not with textual statements.

\begin{table}[!ht]
\centering
\footnotesize
\caption{Evaluation of the effectiveness of numerical representation for ranking purposes. BioBERT models fine-tuned on TREC 2020 and evaluated on CLEF 2020.
}
\begin{tabular}{llccccc}\toprule
& &\multicolumn{4}{c}{CLEF 2020} \\\cmidrule{3-6}
 Model&Represent. &NDCG@10 &P@10 &MRR@10 &MAP \\\midrule
$\mathrm{\emph{CE}}$ (BioBERT) &  &0.2743 &0.2811 &0.4801 &0.1474 \\
\midrule
\multirow{5}{*}{$\textrm{\emph{CE}}_{\textrm{\emph{rel.score}}}$ (dec.)} 
&1 Decimal &0.2263 &0.2261 &0.3322 &0.1064 \\
&2 Decimals &0.2312 &0.2378 &0.3761 &0.1124 \\
&3 Decimals &0.2696 &0.2603 &0.4212 &0.1274 \\
&4 Decimals &\textbf{0.2845} &\textbf{0.2864} &\textbf{0.4852} &\textbf{0.1487} \\
\midrule
\multirow{2}{*}{$\textrm{\emph{CE}}_{\textrm{\emph{rel.score}}}$ (int.)} 
&X100 &0.1897 &0.1974 &0.2453 &0.0742 \\
&X1000 &0.2013 &0.2243 &0.3012 &0.0985 \\
\midrule
$\textrm{\emph{CE}}_{\textrm{\emph{rel.score}}}$ (seg.) & &0.2196 &0.2231 &0.3564 &0.1001 \\
\bottomrule
\label{tab:rep}
\end{tabular}
\end{table}

Results are presented in Table \ref{tab:rep}, which shows that adding floating-point scores performs better than integer and segmented scores, possibly because using floating-point numbers with high precision reduces noise in the input, leading to better performance. While segmented representation provides more information than integer representation, it may cause the model to lose important information about the relative importance of scores.

\paragraph{RQ2: To what extent the effectiveness of the model is affected by the textual information in the statement in addition to the score?}

We experimented with the following statements for \emph{credibility}: 
\begin{enumerate*}
     %\item[$c1.$] \emph{credibility score X}; 
     \item[$c1.$] \emph{Credibility score is X};
     \item[$c2.$] \emph{Credibility score of the document is X}.
\end{enumerate*}
Where $X$ is the credibility score (represented as a 4-decimal floating-point number). We also considered the further addition of \emph{topicality:}
\begin{enumerate*}
     \item[$t1.$] \emph{Topicality score is Y}; 
     \item[$t2.$] \emph{Topicality score of the document is Y}.
\end{enumerate*}
$Y$ is the normalized BM25 score. For the \emph{combination} of topicality and credibility:
\begin{enumerate*}
     \item[$tc.$] $c2$ and $t2$.
 \end{enumerate*} 

\begin{table}[H]\centering
\caption{Evaluation of the effectiveness of statement representation for ranking purposes. BioBERT models fine-tuned on TREC 2020 and evaluated on CLEF 2020.
}\label{tab:state}
\footnotesize
\begin{tabular}{p{2cm}cccccc}\toprule
\multicolumn{2}{c}{} &\multicolumn{4}{c}{CLEF 2020} \\\cmidrule{3-6}Model & Statement&NDCG@10&P@10&MRR@10&MAP \\\midrule
$\textrm{\emph{CE}}_{\textrm{\emph{rel.score}}}$ (dec.) &  &0.2845 &0.2864 &0.4852 &0.1487 \\
\midrule
\multirow{2}{*}{$\textrm{\emph{CE}}_{\textrm{\emph{rel.stat}}}$ (cred., dec.)} 
%&C1 &0.3464 &0.3406 &0.5546 &0.1651 \\
&$c1$ &0.3487 &0.3412 &0.5664 &0.1664 \\
&$c2$ &\textbf{0.3762} &\textbf{0.3669} &\textbf{0.6187} &\textbf{0.1964} \\
\midrule
\multirow{2}{*}{$\textrm{\emph{CE}}_{\textrm{\emph{rel.stat}}}$ (top., BM25)} 
& $t1$ &0.2765 &0.2721 &0.4838 &0.1154 \\
& $t2$ &0.2876 &0.2878 &0.5002 &0.1191 \\
\midrule
$\textrm{\emph{CE}}_{\textrm{\emph{rel.stat}}}$ (cred., dec. and top., BM25)
& $tc$ & 0.3324 &0.331 &0.5576 &0.1745 \\
\bottomrule
\end{tabular}
\end{table}

Results are illustrated in Table \ref{tab:state}. First, they show that adding a statement related to the score -- instead of the score alone -- improves the overall results. Second, enhancing the document with just topicality does not lead to a significant increase in effectiveness. Third, the statement $c2$ related to credibility alone outperforms all models, also the one in which topicality and credibility are considered together.
%Firstly, by comparing Table \ref{tab:rep} and \ref{tab:state } we can observe that adding the statement related to score improves the overall results. \textit{C3} statement outperforms all the models. 
This suggests that the specificity of the statement to the relevance dimension and the document provides valuable information to the model. %, helping it better assess the relevance also with respect to credibility of the document to the query. The experimental results suggest that injecting relevance dimension scores along with the statement in the document can be an effective way to improve the performance of the cross-encoder model when working with health-related documents.

\paragraph{RQ3: What is the effectiveness of the $\textrm{CE}_{\textrm{rel.stat}}$ model compared to baselines?}% combining credibility statement and scores with cross-encoders in the $CE_{\textrm{\emph{rel.stat}}}$ approach compare to other common approaches, and how does it compare to the $CE_{credCAT}$ approach? 
At this point, having selected the best numerical representation for the scores and the best form for the statement, it is possible to answer this research question, which is related to the final evaluation of the proposed model. The results illustrated in Table \ref{tab: final} %demonstrate 
show that the proposed model outperforms other baselines across all four evaluation metrics on both datasets, in particular when the BioBERT textual representation is employed. This result in particular, when also compared with the results illustrated in Tables \ref{tab:rep} and \ref{tab:state}, suggests how enhancing the document with statements related to the relevance dimension is more effective than methods that modify the structure of the cross-encoder input by relevance score injection alone.

\begin{table}[ht]
\caption{Evaluation of the effectiveness of the $\textrm{\emph{CE}}_{\textrm{\emph{rel.stat}}}$ model w.r.t. to baselines. Re-ranking over the first 500 retrieved documents. Results are referred to TREC and CLEF 2020.
}\label{tab: final}
\footnotesize
%\subcaption*{TREC 2020}
\begin{tabular}{llccccc}\toprule
 & &\multicolumn{4}{c}{TREC 2020} \\\cmidrule{3-6}
  Represent. & Model &NDCG@10 &P@10 &MRR@10 &MAP \\\midrule
  \textbf{Lexical} & BM25 &0.4166 &0.4177 &0.5107 &0.2142  \\
  & \emph{WAM} &0.5065 &0.4976 &0.5546 &0.2453 \\
  \midrule
  \textbf{BERT} & $CE_{\textrm{\emph{rel.stat}}}$ &0.6157 &0.5977 &0.7101 &0.3208 \\
  &$CE_{BM25CredCAT}$ &0.5784 &0.5671 &0.6823 &0.2875 \\
  &$CE_{CredCAT}$ &0.5587 &0.5581 &0.6622 &0.2652 \\
  &$CE_{BM25CAT}$ &0.5374 &0.5398 &0.6341 &0.2499 \\
  &$CE$ &0.5589 &0.5501 &0.6619 &0.2664 \\
  \midrule
  \textbf{BioBERT} & $CE_{\textrm{\emph{rel.stat}}}$ &\textbf{0.6704} &\textbf{0.6622} &\textbf{0.7961} &\textbf{0.3865} \\
  &$CE_{BM25CredCAT}$ &0.6219 &0.6245 &0.7512 &0.3324\\
  &$CE_{CredCAT}$ &0.6111 &0.6001 &0.7061 &0.3015  \\
  &$CE_{BM25CAT}$ &0.5875 &0.5812 &0.6801 &0.2765 \\
  &$CE$ &0.6055 &0.6059 &0.6997 &0.2986 \\
\bottomrule
\end{tabular}
\end{table}
%\subcaption*{CLEF 2020}
{\footnotesize
\centering
\begin{tabular}{llccccc}
\toprule
 & &\multicolumn{4}{c}{CLEF 2020} \\\cmidrule{3-6}
  Represent. & Model & NDCG@10 &P@10 &MRR@10 &MAP \\\midrule
  \textbf{Lexical} & BM25 &0.1054 &0.1081 &0.1578 &0.1064   \\
  & \emph{WAM} &0.0865 &0.1002 &0.1232 &0.1102 \\
  \midrule
  \textbf{BERT} & $CE_{\textrm{\emph{rel.stat}}}$ &0.3327 &0.3401 &0.5403 &0.1601\\
  &$CE_{BM25CredCAT}$ &0.3098 &0.3141 &0.5173 &0.1356 \\
  &$CE_{CredCAT}$ &0.2633 &0.2703 &0.4543 &0.1198  \\
  &$CE_{BM25CAT}$ &0.2288 &0.2301 &0.4147 &0.0964 \\
  &$CE$ &0.2579 &0.2601 &0.4456 &0.1165\\
  \midrule
  \textbf{BioBERT} & $CE_{\textrm{\emph{rel.stat}}}$ &\textbf{0.3762} &\textbf{0.3669} &\textbf{0.6187} &\textbf{0.1964} \\
  &$CE_{BM25CredCAT}$ &0.3221 &0.3221 &0.5731 &0.1642 \\
  &$CE_{CredCAT}$ &0.2805 &0.2824 &0.4812 &0.1437   \\
  &$CE_{BM25CAT}$ &0.2414 &0.2522 &0.4702 &0.1274 \\
  &$CE$ &0.2743 &0.2811 &0.4801 &0.1474 \\
  \bottomrule
\end{tabular}
}
\par
% \textbf{RQ3:} To what extent the effectiveness of $CE_{credCAT}$ is dependent on the textual statement words?
% \noindent \par
% We experiment with following textual statements in order to answer this question: 
% \begin{enumerate*}[label=(\roman*)]
%      \item \textit{Credibility-specific:} (\textbf{C1}) `Credibility score X'; \textbf{(C2)} `Credibility score is X'; and \textbf{(C3)} ` Credibility score of the document is X'
%      \item \textit{Topical-specific:} \textbf{(T1)} `Topical score is Y'; and \textbf{(T2)} `Topical score of the document is Y'.
%      \item \textit{Credibility and Topical:}  \textbf{(Comb)} (C3) and (T2)
%  \end{enumerate*} 
% Where X is credibility score (represented as a floating-point number with 4 decimal places) and Y is normalized BM25 score. Results are presented in table \ref{tab:state }.
% Firstly, by comparing Table \ref{tab:rep} and \ref{tab:state } we can observe that adding the statement related to score improves the overall results. \textit{C3} statement outperforms all the models. The possible reason can be the specificity of the statement to the document. Also, it can provide valuable information to the model that helps it better assess the relevance also with respect to credibility of the document to the query. The experimental results suggest that injecting other dimension scores along with the statement in the document can be an effective way to improve the performance of the cross-encoder model when working with health-related documents.

\begin{figure}[H]
  \centering
  \includegraphics[width=\linewidth]{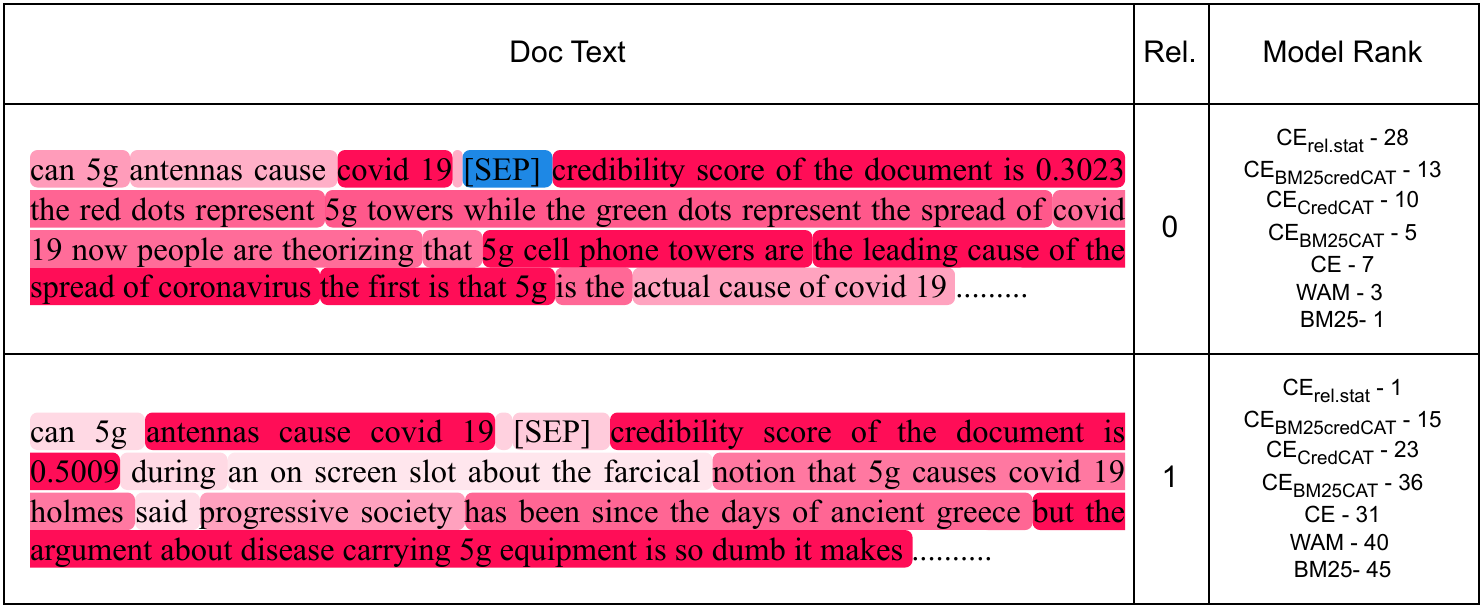}
  \caption{Example query and two passages in the input of $CE_{\textrm{\emph{rel.stat}}}$ highlighted by Shapley values.}
  \label{fig:shap}
\end{figure}

\paragraph{RQ4: Is it possible to explain the contribution that document enhancement makes with respect to ranking effectiveness?}
To provide a reply to this last research question, we computed SHAP (\emph{Shapley Additive Explanations}) \cite{lundberg2017unified} values. The SHAP method allows estimating the importance of each input token, enabling us to understand the model's decision-making process. By applying SHAP to the $CE_{\textrm{\emph{rel.stat}}}$ model, we were able to gain insights into the contribution of the relevance statement and its impact on the model's effectiveness as shown in Figure \ref{fig:shap}. The darker (red) highlighted parts, corresponding to higher SHAP values, are those that contribute most to relevance, both with respect to topicality (the query is represented by the text before [SEP]) and with respect to credibility. We can also see, by analyzing the "Model Rank" column, how our model causes a document deemed irrelevant to fall in the ranking and how it instead places a relevant document at the top.

\section{Conclusion}
\label{sec:conclusion}
In this paper, we introduced the $\mathrm{\emph{CE}}_{\textrm{\emph{rel.stat}}}$ model, which integrates a relevance score and associated textual information to documents in the re-ranking step performed through cross-encoders. Experimental results show that the proposed model statistically significantly outperforms several baselines, either based on the aggregation of multiple relevance scores or cross-encoders that only take topicality into account. The SHAP method used to provide explainability to the results highlighted the contribution of adding relevance statements to documents to increase re-ranking effectiveness.

Future work involves exploring the effectiveness of the %$\mathrm{\emph{CE}}_{\textrm{\emph{rel.stat}}}$ 
model in different domains and with respect to incorporating other relevance dimensions, such as correctness and readability. The use of active learning and other semi-supervised learning methods to improve the efficiency and effectiveness of the model will also be investigated.

\bibliographystyle{ACM-Reference-Format}
\balance
\bibliography{references}
\end{document}